\documentstyle[11pt]{article}

\begin{document}

\begin{flushright}
HU--TFT--93--29\footnote[1]
{Submitted for publication in Phys. Lett. B}\\
May 1993
\end{flushright}
\bigskip
\begin{center}
{\Large\bf
Abelian current algebra and the Virasoro algebra\\
\medskip on the lattice}\\
\bigskip
{\large\bf
     L. Faddeev$^{1,}$\footnote[2]
     {Supported by the
 Russian Academy of Sciences and the Academy of Finland}
     and A. Yu. Volkov$^{2,}$\footnote[3]
     {On leave of absence from
Saint Petersburg Branch of the Steklov Mathematical Institute,\\
Fontanka 27, Saint Petersburg 191011, Russia}}
\end{center}
\bigskip
\begin{quote}
$^1$ {\em Saint Petersburg Branch of the Steklov Mathematical Institute\\
     Fontanka 27, Saint Petersburg 191011, Russia\\{\em and}
     Research Institute for Theoretical Physics\\
 P. O. Box 9 (Siltavuorenpenger 20C), SF-00014
 University of Helsinki, Finland\\
$^2$ Physique - Math\'{e}matique, Universit\'{e} de Bourgogne\\
     B.P.138 Dijon C\'{e}dex, France}

\bigskip
{\bf Abstract.} We describe how a natural lattice analogue of the
 abelian current algebra combined with free discrete time dynamics
 gives rise to the lattice Virasoro algebra and corresponding hierarchy
 of conservation laws.
\end{quote}

\section*{Introduction}

Two modern directions in 1+1-dimensional quantum field
theory -- we mean conformal field theory (CFT) and quantum inverse
scattering method -- that developed at first quite independently have
eventually proved to be closely related. The following observation
\cite{G}, for instance, provides a clear evidence of their
connection. The central place in CFT belongs to the Virasoro algebra
whose generators $L_m$ are the Fourier modes of the improved
stress--energy tensor having before quantization the Poisson bracket
$$ \{t(x),t(y)\}=2\gamma (t(x)+t(y))\delta^\prime (x-y)+
            \gamma\delta^{\prime \prime \prime}(x-y)\;\;\; .      $$
This very bracket is well known in the inverse scattering method as
one of the brackets in the KdV hierarchy of Poisson structures.
Combined with the similar observation concerning $W$--algebras it
provides a good reason to think that integrable models are associated
somehow with conformally invariant ones or, if you wish, are their
deformations.

While this idea is quite transparent for the classical equations of
motion and Poisson brackets, the corresponding quantum picture is rather
patchy. On one hand, an approach proposed in \cite{FF} allows in
principle to construct quantum commuting conservation laws for the
Virasoro algebra. On the other hand, neither \cite{FF} nor the quantum
inverse scattering method deliver reasonably
effective construction of that laws. We hope that recent expansion
of the latter method to the models in discrete space--time
\cite{DdV,N,FV92} will possibly throw new light on the nature of quantum
integrability in general and the structure of conservation laws
in particular.

At the same time, since the discrete space--time picture happened to be
quite effective for the sine--Gordon model it would be natural to
try to discretize also CFT. This idea is not completely new. Our approach
here differs from the previous efforts \cite{G,FT,V88,B,V92} in
discretizing not only the spatial but also the time variable.

We will consider the simplest model where the above mentioned
stress--energy tensor emerges as the Miura transformation
$$  t=p^2+p^{\prime}  $$
of the periodic free field (abelian current) $p(x)$ having the Poisson
bracket
$$  \{ p(x),p(y) \}=\gamma \delta ^\prime (x-y)   $$
and the equation of motion
$$  p_t(x,t)+p_x(x,t)=0  $$
$$  p(x,0)=p(x)\;\;\;.        $$

In section 1 we introduce the corresponding lattice model.
In section 2 we examine its evolution operator.
In section 3 we find lattice counterparts of the Miura transformation
and the improved stress--energy tensor. In section 4 we describe a
family of commuting operators which contains this evolution
operator. In section 5 we introduce a suitable exchange algebra
and modify accordingly the evolution operator.

\section{Current algebra on the lattice
          and the free field model in the discrete space--time}

Consider the operators $w_n , n=1,\ldots ,2N$ with the commutation
relations
$$  w_{n-1} w_n = q^2 w_n w_{n-1} \;\;\;,\;\;\;n=2,3,\ldots ,2N $$
$$  w_{2N} w_1 = q^2 w_1 w_{2N}   $$
$$  w_m w_n = w_n w_m\;\;\;{\rm if}\;\;\;1<|m-n|<2N-1\;\;\;, $$
where $q$ is a complex constant. Adopting the notations
$$  w_{n+2kN}=w_n\;\;\;,\;\;\;n=1,2,\ldots ,2N\;\;\;k\in{\bf Z}  $$
and introducing the periodic Kroneker symbol
$$  \delta_n\equiv\sum_{k\in{\bf Z}}\delta_{n,2kN}    $$
one can collect all commutation relations into a single formula
$$  w_m w_n = q^{2(\delta_{m-n+1}-\delta_{m-n-1})} w_n w_m
    \;\;\;,\;\;\;m,n\in{\bf Z}\;\;\;.      $$

In the classical limit $q=e^{i\hbar\gamma},
\hbar\rightarrow 0$ this leads to the Poisson bracket
$$ \{ w_m,w_n \}
= -2\gamma(\delta_{m-n+1}-\delta_{m-n-1})w_m w_n\;\;\;.$$
In the continuous limit, setting $x=n\Delta,N\Delta=2\pi,
w_n \sim e^{2\Delta p(x)}$ and $\Delta \rightarrow 0$
one gets precisely the
periodic free field mentioned in the introduction.

Unfortunately, the most desirable case $|q|=1$ is rather difficult and
we prefer to start with less realistic but simpler one
with $q$ inside the
unit circle. We will also temporarily adopt one more simplifying
assumption. Up to the final section the two central elements of
the algebra of $w$'s
$$  C_1=\prod w_{odd}   $$
$$  C_2=\prod w_{even}  $$
will always assumed be equal
$$   C_1=C_2\;\;\;.    $$

We have completed the description of the ``phase space" and are now to
choose the ``equations of motion". As the spatial variable is already
discretized, it is only natural to make the time to be discrete as well.
Then the inevitable replacement for the continuous free
equation $ p_t=-p_x $ will be just
$$  w_n (t+1)=w_{n-1} (t)\;\;\;,\;\;\;t\in {\bf Z}  $$
$$  w_n (0)=w_n \;\;\; .      $$
As such a dynamics preserves commutation relations between $w$'s, there
should exist an ``evolution" operator $U$ such that
$$  w_n (t+1) = U^{-1} w_n (t) U \;\;\; . $$
Comparing the last two formulas one sees that $U$ is the shift operator
$$  w_n U=U w_{n-1}   $$
which will be the subject of our interest in the rest of this paper.

\section{Shift operator and the braid group}

In this section we will show that the operator
$$  U=h_1 h_2 \ldots h_{2N-1} $$
(notice that the product is one site shorter than the lattice) is indeed
the shift operator if
$$  h_n=\theta(w_n)  $$
$$  \theta(w)=\sum_{k\in{\bf Z}}q^{k^2}w^k=\theta_0 (q,w)\;\;\;,   $$
where $\theta_0 (q,w)$ is the theta--function.

To verify this statement let us note first that the theta--function
satisfies the functional equation
$$   \frac{\theta(qw)}{\theta(q^{-1}w)}=\frac{1}{w} \;\;\;.   $$
This allows to obtain commutation relations for the operators $w_n$
and $h_n$ , for instance,
$$  w_n h_{n-1}=w_n \theta(w_{n-1})=\theta(q^{-2}w_{n-1})w_n=
  q^{-1}\theta(w_{n-1})w_{n-1}w_n=q^{-1}h_{n-1}w_{n-1}w_n \;\;\; . $$
Now one easily obtains the crucial commutation relation
$$  w_n h_{n-1} h_n = h_{n-1} h_n w_{n-1}   $$
which for $n=2,3,\ldots,2N-1$ yields immediately
$$  w_n U = \ldots w_n h_{n-1} h_n \ldots
= \ldots h_{n-1} h_n w_{n-1} \ldots=
             U w_{n-1}\;\;\; .  $$
Due to the existence of the two central elements in the algebra of $w$'s
the remaining two relations are determined by those $2N-2$ which we
have already got:
$$  w_1 U = \frac {C_1}{w_3 w_5 \ldots w_{2N-1}} U =
   U \frac {C_1}{w_2 w_4 \ldots w_{2N-2}}= U \frac{C_1}{C_2}w_{2N}  $$
$$  w_{2N}U=U \frac{C_2}{C_1}w_{2N-1}\;\;\; .   $$
Since $C_1=C_2$,
this completes the proof that $U$ is the shift operator, i.e.
that for any $n$ from $1$ to $2N$
$$   w_n U = U w_{n-1} \;\;\;.  $$

Several remarks are in order now. First, the commutation relation
between $w$ and the pair of $h$'s is valid not only for
bare $w$ but for any function of it and for instance
for $\theta(w)$ itself. This means that the
operators $h_n$ generate the braid group of the $A_{2N-1}^{(1)}$ type:
$$  h_n h_{n-1} h_n
= h_{n-1} h_n h_{n-1}\;\;\;,\;\;\;n=2,3,\ldots ,2N $$
$$  h_1 h_{2N} h_1 = h_{2N} h_1 h_{2N}\;\;\;.  $$

Second, the proof and the definition of the operator
$U$ may create the impression that $2N$-th site of the lattice is
somehow distinguished. That is not so just because
$$   U = h_n h_{n+1} \ldots h_{n+2N-1}    $$
for any $n\in{\bf Z}$.

Third, the possibility to express the shift operator as a
product of local factors certainly reflects the fact that
the corresponding
continuous dynamics $p_t=-p_x$ is generated by the local hamiltonian
$$    H=\frac{1}{2\gamma}\int_0^{2\pi} p^2(x){\rm d}x\;\;\;.    $$
And however different the theta--function and the density
of this hamiltonian $p^2$ may look, the well known formula
$$    \theta_0 (q,q^{2n})=q^{-n^2}\theta_0 (q,1)    $$
shows that there is no contradiction between them.

\section{Miura transformation and the Virasoro algebra}

In the continuous theory the Miura transformation $t=p^2+p^{\prime}$
turned the free field into the improved stress--energy tensor or, in
the hamiltonian language, into the ``improved" hamiltonian density
$$   H=\frac{1}{2\gamma}\int_0^{2\pi} p^2(x){\rm d}x
   =\frac{1}{2\gamma}\int_0^{2\pi} (p^2(x)+p^{\prime}(x)){\rm d}x
     =\frac{1}{2\gamma}\int_0^{2\pi} t(x){\rm d}x\;\;\;.$$
We will now describe a similar effect on the lattice. More precisely,
we shall show that
$$   U=h_1 h_2 \ldots h_{2N-1}       $$
$$ \approx s(w_1)\left( s(w_1^{-1})s(w_2)s(w_2^{-1})\ldots
                  s(w_{2N-1})s(w_{2N-1}^{-1})\right)
\;\;\;\;\;\;\;\;\;\;\;\;$$  $$\;\;\;\;\;\;\;\;\;\;
=\left( s(w_1^{-1})s(w_2)s(w_2^{-1})s(w_3)\;\;\ldots\;\;
              s(w_{2N-1}^{-1})\right) s(w_{2N})  $$
$$    =s(w_1^{-1}+w_2+qw_1^{-1}w_2)s(w_2^{-1}+w_3+qw_2^{-1}w_3)\ldots
       s(w_{2N-1}^{-1}+w_{2N}+qw_{2N-1}^{-1}w_{2N}) \;\;\;  ,    $$
where the function $s(w)$ has the form
$$ s(w)=1+\sum_{k=1}^\infty \frac{q^{\frac {k(k-1)}{2}}}
            {(q^{-1}-q)(q^{-2}-q^2)\ldots (q^{-k}-q^k)}w^k   $$
and the sign $\approx$ means the equality up to a constant factor.

One clearly sees that for this chain of transformations to be
possible the function $s(w)$ should have two properties:
\begin{description}
\item[1.]$\;\;\;s(w)s(w^{-1}) \approx \theta(w)\;\;\;;$
\item[2.]$\;\;\;s(v)s(u)=s(v+u+qvu)
\;\;\;$if $u,v$ is a Weyl pair, i.e. $uv=q^2vu\;\;\;$.
\item It indeed possesses both ones and the list of its notable
features can be continued:
\item[3.]$\;\;\;s(u)s(v)=s(u+v)\;\;\;$if again$\;\;\;uv=q^2vu\;\;\;.$
\end{description}

All three equalities may be verified by the straightforward but elaborate
calculation using the explicit formulas for the both functions involved.
We prefer however to do without all that combinatorics. The technics
which we are going to outline below will be based on the observation
that the function $s(w)$ satisfies the functional equation
$$   \frac{s(qw)}{s(q^{-1}w)}=\frac{1}{1+w}\;\;\;.    $$

It is now easy to see why the first of the three equalities is true.
Indeed, both sides of it satisfy the same functional equation
$$    \frac{s(qw)s((qw)^{-1})}{s(q^{-1}w)s((q^{-1}w)^{-1})}=\frac{1}{w}
      =\frac{\theta(qw)}{\theta(q^{-1}w)}                 $$
and thus the functions $s(w)s(w^{-1})$ and $\theta(w)$ coincide
(up to the constant $s^2(1)/\theta(1)$) at the
points $w=q^{2k},k\in {\bf Z}$. This in fact means that they
coincide everywhere. To make such a conclusion one certainly
has to examine first the analitycal properties of that functions.
We leave this in many ways important question to be discussed
in detail elsewhere.

To verify the second equality let us first note that the operators
$$  {\cal U}=u+quv^{-1}   $$
$$  {\cal V}=u+v+qvu      $$
form just another Weyl pair
$$  {\cal UV} = q^2 {\cal VU}     $$
which is ``dual" to the original one in the sense that the inverse map
$$  u^{-1}={\cal U}^{-1}+{\cal V}^{-1}+q{\cal V}^{-1}{\cal U}^{-1} $$
$$  v^{-1}={\cal V}^{-1}+q{\cal U V}^{-1}               $$
is similar to the direct one. The r.h.s. $s({\cal V})$ of the equality
which we are verifying satisfies the commutation relations
$$  [s({\cal V}),{\cal V}\;]=0   $$
$${\cal U} s({\cal V}){\cal U}^{-1} =
\frac{s({\cal V})}{1+q{\cal V}}\;\;\;. $$
One may check that the same is true for the l.h.s. $s(v)s(u)$. This
means (at least on the formal level adopted here) that $s({\cal V})$
and $s(v)s(u)$ coinside up to a constant which is equal to $1$
just because $s(0)=1$.

The third property of $s$ we actually do not need and will not
verify (this could be done in roughly the same manner as for
the second one). Note that it is the standard functional relation
for the so called $q$--exponential.

So, we have at last established that
$$   U=s_1 s_2 \ldots s_{2N-1}    $$
where
$$   s_n=s(w_n^{-1}+w_{n+1}+qw_n^{-1}w_{n+1})\;\;\; .  $$
The appearance of the combinations of the operators $w_n$ involved here
is hardly surprising.
As was gradually understood during the investigation of the
Liouville equation on the lattice \cite{FT,V88,V92,B,F} the discrete
Miura transformation is likely to have the form
$$t_n
={\textstyle \frac{1}{4}}(1+w_n^{-1}+w_{n+1}+qw_n^{-1}w_{n+1})\;\;\;.$$
The ``stress--energy" tensor $t_n$ defined like this has the correct
continuous classical limit
$$  t_n\sim 1-\Delta^2 t(x)     $$
as well as closed commutation relations
and some other promising properties \cite{V92,B,F}.
That gives the right to say that $t$'s generate the lattice
Virasoro algebra. It was not however easy to
get real benefits of this construction because on the lattice neither a
suitable Fourier transform nor the ``diffeomorphisms" are available yet
and thus one cannot obtain lattice counterparts of the
generators $L_m$ of the Virasoro algebra directly from $t$'s.
We have just made the first step in this direction and established
in what sense the ``generator" of the only evident lattice
``diffeomorphism" is the zero'th ``Fourier mode" of the discrete
stress--energy tensor. At the same time we seem to get the key to the
understanding of the whole Fourier transform relevant for our discrete
free field theory.

\section{Yang--Baxter equation}

We now change the direction and turn from the ``conformal" side of our
discrete--discrete free field model to its link with the quantum
inverse scattering method.

Another name for the braid group commutation relation (see sect.2)
is the Yang--Baxter equation. In this context it reads:
$$ \theta (v) \theta (u) \theta (v) = \theta (u) \theta (v) \theta (u)
\;\;\;{\rm if}\;\;\;
   uv=q^2vu\;\;\;  .  $$
The argument of the theta--function is not to be mistaken here
for a spectral parameter.
Genuine Yang--Baxter equation with (multiplicative) spectral
parameter should have the form:
$$ r(\lambda ,v) r(\lambda \mu ,u) r(\mu ,v)
   =r(\mu ,u) r(\lambda \mu ,v) r(\lambda ,u)
\;\;\;{\rm if}\;\;\;
   uv=q^2vu\;\;\;  ,  $$
the ``$R$-matrix" $r(\cdot,\cdot)$ here being a function of two
variables first of which is a complex number and the second one is an
operator.
A suitable for us solution to this equation has been found in \cite{FZ}:
$$  r(\lambda,w)=1+
      \sum_{k=1}^\infty \frac
      {(1-\lambda)(q-\lambda q^{-1})\ldots (q^{k-1}-\lambda q^{-k+1})}
     {(q^{-1}-\lambda q)(q^{-2}-\lambda q^2)\ldots (q^{-k}-\lambda q^k)}
                                                (w^k+w^{-k})\;\;\;.$$
It is easy to notice that
$$  r(0,w)=\theta (w)  $$
and thus expect the function $r(\lambda ,w)$ to satisfy some functional
equation similar to the one satisfied by the theta--function but with
the deformed r.h.s.. Such an equation was proposed in \cite{V92}:
$$   \frac{r(\lambda,qw)}{r(\lambda,q^{-1}w)}
=\frac{1+\lambda w}{\lambda +w}\;\;\;. $$
The natural order of things is probably opposite to the historical one
presented above. Both the explicit form of the function $r$ and the
fact that it satisfies the Yang--Baxter equation are actually
the corollaries of this functional equation.
It may be also useful to know that with respect to its another argument
the function $r(\lambda,w)$ satisfies the equation
$$\frac {r(q\lambda,w)}{r(q^{-1}\lambda,w)}
=(1+\lambda w)(1+\lambda w^{-1})\;\;\;.$$

Let us forget for a while about our finite periodic lattice and
consider instead an infinite one with a suitable rapidly decreasing
boundary conditions for the operators $w_n,n\in {\bf Z}$.
Multiplying the R-matrices $r(\lambda ,w_n)$ along the lattice
$$ U(\lambda)
=\ldots r(\lambda ,w_{n-1})r(\lambda ,w_n)r(\lambda ,w_{n+1})\ldots  $$
we obtain the commuting family of operators
$$    [U(\lambda),U(\mu)]=0    $$
containing the shift operator
$$     U=U(0)\;\;\;  .     $$
Following the guidelines worked out in the previous section we
factorize the function $r(\lambda,w)$
$$  r(\lambda,w)\approx s(\lambda,w)s(\lambda,w^{-1})  $$
$$  s(\lambda,w)=\frac{s(w)}{s(\lambda w)}=1+
      \sum_{k=1}^\infty \frac
      {(1-\lambda)(q-\lambda q^{-1})\ldots (q^{k-1}-\lambda q^{-k+1})}
      {(q^{-1}-q)(q^{-2}-q^2)\ldots (q^{-k}-q^k)}w^k  $$
$$   \frac{s(\lambda,qw)}{s(\lambda,q^{-1}w)}
=\frac{1+\lambda w}{1+w}\;\;\;,
  \;\;\;\frac{s(q\lambda,w)}{s(q^{-1}\lambda,w)}=1+\lambda w\;\;\;, $$
then discover that the multiplication rule still holds
$$   s(\lambda,v)s(\lambda,u)
=s(\lambda,v+u+qvu)\;\;\;,\;\;\;uv=q^2vu\;\;\;,  $$
then rearrange accordingly the decomposition of $U(\lambda)$
$$  U(\lambda)
\approx\ldots s_{n-1}(\lambda) s_n (\lambda) s_{n+1}(\lambda)\ldots $$
$$   s_n(\lambda)=s(\lambda,w_n^{-1}+w_{n+1}+qw_n^{-1}w_{n+1})   $$
and eventually find out that $U(\lambda)$ is the quantum lattice
counterpart of the generating function of the KdV conservation
laws hierarchy.

Unfortunately, the periodic case is harder to control and the natural
and basically correct scheme outlined above is not
easy to apply there (we will address this problem in the forthcoming
paper). Nevertheless, the emergence of a fully fledged $R$--matrix
in the free lattice theory indicates that the link between conformal
invariance and complete integrability has not disappeared on the way
to the lattice. It also comes as no surprise that the same $R$--matrix
happens to be heavily involved in the construction of the evolution
operator and the conservation laws of the quantized
Hirota (discrete--discrete sine--Gordon) model \cite{FV92}.
All that just firms our belief that good
lattice models inherit some really essential properties of their
continuous counterparts being at the same time much simpler to live with
and thus providing a good starting point for the investigation of
1+1-dimensional quantum field theories.

\section{Exchange algebra}

Let us come back to the more general case when $C_1\neq C_2$
(see sect.1). Notice first of all that the shift operator which we
discussed up to now would permute $C_1$ and $C_2$ :
$$ C_1 U=U C_2\;\;\;,\;\;\;C_2 U=U C_1\;\;\;.   $$
It only means that if they are not equal this operator does not exist.
What does exist in that case is the operator shifting by two sites:
$$  w_n V = V w_{n-2}\;\;\;. $$
Its local decomposition has certainly the same density as that of the
operator $U$ but is two times longer:
$$V=h_2h_3\ldots h_{2N}h_1h_2\ldots h_{2N-1}
=h_nh_{n+1}\ldots h_{n+4N-2}\;\;\;.$$
All that we did with the operator $U$ can be now almost literally
repeated for $W$. At the same time we gain the possibility to realize
the algebra of the operators $w_n$ as a subalgebra of the so called
exchange algebra.

To make it clear let us first turn to the classical continuous case
where that algebra has the form
$$  \{\psi(x),\psi(y)\}=-{\textstyle \frac{1}{2}}\gamma\epsilon(x-y)
         \psi(x)\psi(y)\;\;\;,\;\;\;0<x,y<2\pi\;\;\;,   $$
where $\epsilon(x)$ is the sign function. This is certainly only a
half of the definition. It remains to specify the boundary condition.
Let us introduce the quasimomentum $C$
$$  \{C,\psi(x)\}=-\gamma C \psi(x)    $$
and continue the field $\psi(x)$ outside the fundamental domain
by means of the quasiperiodicity condition
$$  \psi(x+2\pi)=C\psi(x)\;\;\; .  $$
As a result we get the Poisson bracket on the whole real axis
$$ \{\psi(x),\psi(y)\}=
 -{\textstyle \frac{1}{2}}\gamma\epsilon_{2\pi}(x-y)\psi(x)\psi(y)
 \;\;\;,$$
where the function $\epsilon_Q (x)$ is a kind of quasiperiodic
sign function:
$$ \epsilon_{2\pi}(x)=2k+1\;\;\;{\rm if}\;\;\;
              2\pi k<x<2\pi(k+1)\;\;\;,\;\;\;k\in{\bf Z}\;\;\;.$$
Taking the logarithmic derivative of $\psi(x)$ we obtain the
familiar periodic free field
$$  p=\frac{\psi^\prime}{\psi}\;\;\;,\;\;\;p(x+2\pi)=p(x)  $$
$$  \{ p(x),p(y) \}=\gamma \delta ^\prime (x-y)\;\;\;.  $$
The quasimomentum $C$ becomes the central element
$$  C=\exp (\int_0^{2\pi} p(x){\rm d}x)    $$
$$  \{C,p(x)\}=0 \;\;\; .    $$
It is needless to say that
the real reason for the introduction of the exchange algebra is
its remarkable connection with the Virasoro algebra via the
Sturm--Liouville equation:
$$   -\psi^{\prime \prime}+t\psi=0 \;\;\;,   $$
where $t$ is the stress-energy tensor $t=p^2+p^{\prime}$ (see
Introduction).

Analogous construction on the lattice is a bit more sophisticated.
We introduce $2N+2$ operators
$$  \psi_1,\psi_2,\ldots,\psi_{2N}\;\;\;{\rm and}\;\;\;C_1,C_2   $$
with the commutation relations
$$  \psi_{2m} \psi_{2n+1} = q \psi_{2n+1} \psi_{2m}\;\;\;{\rm if}\;\;\;
    1<2m<2n+1<2N           $$
$$  \psi_{2m+1} \psi_{2n} = q \psi_{2n} \psi_{2m+1}\;\;\;{\rm if}\;\;\;
    1\leq 2m+1<2n\leq 2N           $$
$$  \psi_{2n+1}C_1=q^2C_1\psi_{2n+1}   $$
$$  \psi_{2n}C_2=q^2C_2\psi_{2n}   $$
$$  [\psi_{2m+1},\psi_{2n+1}]=[\psi_{2m},\psi_{2n}]=[\psi_{2m+1},C_2]=
    [\psi_{2m},C_1]=[C_1,C_2]=0\;\;\;  .   $$
Defining then $\psi_n$ for the values of $n$ outside
the 1 to $2N$ range by
$$ \psi_{2n+1+2N}=C_2\psi_{2n+1}
\;\;\;,\;\;\;\psi_{2n+2N}=C_1\psi_{2n}\;\;\;,  $$
we collect all commutation relations into a single formula
$$  \psi_m \psi_n=q^{-\epsilon_{m-n}}\psi_n \psi_m\;\;\;,  $$
where
$$  \epsilon_{2n}=0 \;\;\;,\;\;\;\epsilon_{2n+1}=2k+1
\;\;\;{\rm if}\;\;\;
    2kN\leq 2n+1 < 2(k+1)N\;\;\;.  $$
Taking the ``derivatives" of $\psi$'s
$$  w_n=\frac{\psi_{n+1}}{\psi_{n-1}}    $$
we reproduce precisely the lattice current algebra introduced in sect.1.

``In between" of these exchange algebra of $\psi$'s and current algebra
of $w$'s producing the models with $N+1$ and $N-1$ degrees of freedom
lies a couple of useful algebras both giving models
with $N$ degrees of freedom.
The first one is generated by the operators
$$  \phi_1,\phi_2,\ldots,\phi_{2N}\;\;\;{\rm and}\;\;\;C\;\;\;,  $$
where
$$  \phi_n=\psi_n\psi_{n+1}\;\;\;,\;\;\;C=C_1C_2  $$
and thus
$$    \phi_{n+2N}=C\phi_n \;\;\; .      $$
The commutation relations are
$$   \phi_m \phi_n = q^{-2\varepsilon_{m-n}}\phi_n \phi_m   $$
$$   \varepsilon_{2kN}=2k\;\;\;,\;\;\;
\varepsilon_n=2k+1\;\;\;{\rm if}\;\;\;
   2kN<n<2(k+1)N  $$
and there is a central element
$$   c=\frac{C_1}{C_2}=\frac{\prod_{n=1}^{N} \phi_{2n-1}^{-1}\phi_{2n}}
    {\prod_{n=1}^{N} \phi_{2n}^{-1}\phi_{2n+1}}\;\;\;. $$
The second subalgebra is generated by the operators
$$ \psi_2,\psi_4,\ldots,\psi_{2N}\;\;\;{\rm and}\;\;\;
w_2,w_4,\ldots,w_{2N}$$
with extremely simple commutation relations
$$   \psi_{2n}w_{2n}=q^2w_{2n}\psi_{2n}   $$
$$   [\psi_{2m},w_{2n}]=0\;\;\;{\rm if}\;\;\;m\neq n   $$
$$   [\psi_{2m},\psi_{2n}]=[w_{2m},w_{2n}]=0\;\;\;.   $$
First of them looks quite natural and is not bad at all on the infinite
lattice. The second one consists just of the independent
Weyl pairs and proves to be rather helpful in some
applications \cite{FV92}.

To cover all emerging versions of the
lattice free field model we need to
find the shift operator which is good
for the largest of these algebras:
$$ \psi_n W
=W \psi_{n-2}\;\;\;{\rm for\;any}\;\;n\in{\bf Z} \;\;\; .   $$
The previous shift operator $V$ is certainly almost correct
and just one more factor turns it into $W$:
$$   W=\frac{V}{\theta(\frac{C_1}{C_2})}
    =\frac
  {\theta(\frac{\psi_{n+1}}{\psi_{n-1}})\theta(\frac{\psi_{n+2}}
   {\psi_{n}})
  \ldots \theta(\frac{\psi_{n+4N-1}}{\psi_{n+4N-3}})}
             {\theta(\frac{C_1}{C_2})}\;\;\; .  $$
\section*{Conclusion}

The $U(1)$ current algebra obviously corresponds to the case
when $|q|=1$ and the operators $w_n$ are unitary.
The series which defined
the functions $\theta(w),r(\lambda,w),s(w),s(\lambda,w)$ do
not make much sense when $|q|=1$, especially when $q$ is not a
root of unity. Fortunately, when $q$ is a root of unity
$$  q^l=1\;\;\;,  $$
and the operators $w_n$ are normalized by the
same condition
$$  w_n^l=1  $$
the equations
$$   \frac{\theta(qw)}{\theta(q^{-1}w)}=\frac{1}{w}     $$
$$   \frac{r(\lambda,qw)}{r(\lambda,q^{-1}w)}
=\frac{1+\lambda w}{\lambda +w}  $$
still can be solved:
$$  \theta(w)=\sum_{k=0}^{l-1}q^{k^2}w^k   $$
$$  r(\lambda,w)=1+
      \sum_{k=1}^{l-1} \frac
      {(1-\lambda)(q-\lambda q^{-1})\ldots (q^{k-1}-\lambda q^{-k+1})}
{(q^{-1}-\lambda q)(q^{-2}-\lambda q^2)
 \ldots (q^{-k}-\lambda q^k)}w^k\;\;\;.  $$
Unfortunately, the same cannot be said about the equation
$$\frac{s(\lambda,qw)}{s(\lambda,q^{-1}w)}=\frac{1+\lambda w}{1+w}  $$
which has the solution only at the points
$$    \lambda=q^{2k}  \;\;\; .  $$
The case thus deserves separate investigation which is now in progress.

One could compile quite long list of
other questions left here without answers.
How to turn formal operator functions
and calculations into something real, what
is the precise definition of the commuting family $U(\lambda)$
in the periodic case, how to extract from there
not only the shift operator but something
like a hierarchy of local conservation
laws, what are the "Fourier modes" of the discrete stress--energy
and "diffeomorphisms" of the discrete circle?
We will address some of these
questions in forthcoming papers.
However, we still did not mention what is
probably the most interesting
problem in this context. The current algebra considered in the paper
is abelian. It is natural to ask whether one can apply similar technics
to a model where the phase space is a nonabelian
current algebra and the equations of
motion are again free, i.e. to the chiral
WZNW model. A good example of the lattice $SL(N)$ current algebra
has been constructed in \cite{RS,AFSV}. Since in the $SL(2)$
case we know a quite convenient free field parametrization of the
lattice currents, the corresponding discrete--discrete WZNW model can
be treated more or less in the same
way as the abelian free field \cite{FVP}. It would be however more
interesting to develop an essentially nonabelian formalism
not appealing to free fields. We can't help feeling that some of the
constructions in this paper are more general than they look and
thus their nonabelian generalisation is quite possible.

\bigskip
\noindent{\bf Acknowledgement}\\
We are grateful to B. Feigin, A. Kirillov, E. Sklyanin
and M. Semenov-Tian-Shansky for discussions. L. Faddeev is grateful
to professors M. Flato and P. Mitter
for their hospitality during his visits to Universit\'{e} de Bourgogne
and Universit\'{e} Paris VI where this work has been completed.

\end{document}